%% file: main.tex
\documentclass[journal]{IEEEtran}

\usepackage[pdfborder={0 0 0},colorlinks=false,bookmarksopenlevel=2,pdfstartview=Fit,pdfview=FitH]{hyperref}

\usepackage{graphicx}

\usepackage{subfigure}
\usepackage{layouts}
\usepackage{multicol, blindtext}
\usepackage[colorinlistoftodos]{todonotes}
\setuptodonotes{inline}

\usepackage{siunitx}

\ifCLASSINFOpdf
\else
\fi

\newcommand{\cvisi}{\text{CV}_{\text{ISI}}}

\begin{document}
\title{
Characterization of Off-wafer Pulse Communication in BrainScaleS Neuromorphic System
}

\author{
\IEEEauthorblockN{
    Bernhard~Vogginger$^\dagger$,
    Vasilis~Thanasoulis$^\dagger$,
	Johannes~Partzsch,
    Christian~Mayr
}\\
\IEEEauthorblockA{
\textit{Chair of Highly-Parallel VLSI-Systems and Neuro-Microelectronics} \\
\textit{Technische Universität Dresden},
Dresden, Germany \\
Email: bernhard.vogginger@tu-dresden.de}

\thanks{The authors marked with $\dagger$ contributed equally to this work.}%
}

\maketitle

\begin{abstract}

Neuromorphic VLSI systems take inspiration from biology to enable efficient emulation of large-scale spiking neural networks and to explore new computational paradigms.
To establish large neuromorphic systems, a sophisticated routing infrastructure is needed to communicate spikes between chips and to/from the host computer.
For the BrainScaleS wafer-scale neuromorphic system considered in this work, especially the stimulation with input spikes and the recording of spikes is demanding, requiring high bandwidth and temporal resolution due to the accelerated emulation of neural dynamics 10.000 faster than biological real time.
Here, we present a systematic characterization of the BrainScaleS off-wafer communication infrastructure implemented around Kintex7 FPGAs.
The communication flow is characterized in terms of throughput, transmission delay, jitter and pulse loss.
Further, we analyze the effect of the communication distortions (like pulse loss and jitter) on a neural benchmark model with highly varying spike activity.
The presented methods and techniques for communication evaluation are general applicable and provide useful insights for the mapping of network models to the hardware such as the distribution of input spikes across communication channels.

\end{abstract}


\IEEEpeerreviewmaketitle

\section{Introduction}
\label{sec_intro}

\input{01_Introduction}

\section{Material and Problem Statement}
\label{sec_material}
\input{02_Material_and_Problem_Statement}

\section{Methods}  
\label{sec_methods}

\input{03_Methods}

\section{Characterisation of the Communication}  
\label{sec_characterisation}
\input{04_Characterisation_of_the_Communication}

\section{Neuromorphic Benchmark Application}
\label{sec_benchmark}

\input{05_Benchmark_Application}

\section{Discussion}
\label{sec_results}
\input{06_Discussion}

\section{Conclusion}
\label{sec_conclusion}

\input{07_Conclusion}

\section*{Acknowledgment}
This research work was funded from the European Community's Seventh Framework Programme under grant agreements no 237955 (FACETS-ITN), 269921 (BrainScaleS) by the EC Horizon 2020 Framework Programme under grant agreements 720270 (HBP SGA1), 785907 (HBP SGA2).
We thank our colleagues Stefan Schiefer and Stephan Hartmann for their help with the FPGA boards and the Electronic Vision(s) from Heidelberg University for the collaboration on the BrainScaleS system.

\ifCLASSOPTIONcaptionsoff
  \newpage
\fi

\bibliographystyle{IEEEtran}
\bibliography{references}

\end{document}

%% file: 01_Introduction.tex
\begin{figure*}[!ht]
\centering
\includegraphics[width=0.9\textwidth]{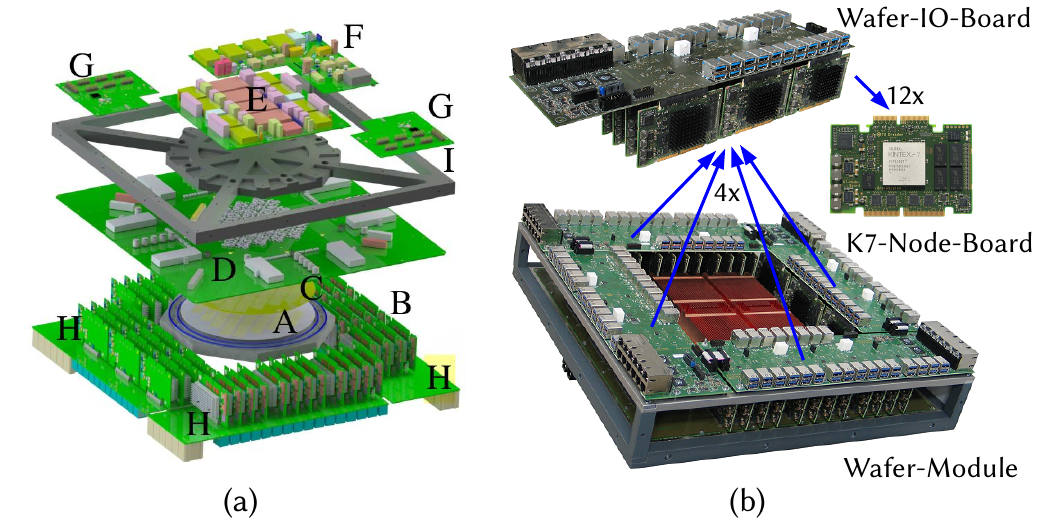}
\caption{BrainScaleS system:
(a) 3D-schematic of a BrainScaleS wafer module (dimensions: 50 cm
× 50 cm × 15 cm) hosting the wafer (A) and 48 FPGAs (B). The positioning
mask (C) is used to align elastomeric connectors that link the wafer to the
large main PCB (D). Support PCBs provide power supply (E \& F) for the
on-wafer circuits as well as access (G) to analog dynamic variables such
as neuron membrane voltages. The connectors for inter-wafer (USB slots)
and off-wafer/host connectivity (Gigabit-Ethernet slots) are distributed over
all four edges (H) of the main PCB. Mechanical stability is provided by an
aluminum frame (I). © [2017] IEEE. Reprinted with permission from \cite{schmitt2017neuromorphic}. 
(b) Photograph of fully assembled wafer module with wafer-IO-board and Kintex-7 FPGA board.
}
\label{fig_system}
\end{figure*}

\IEEEPARstart{I}{n} recent years there is a particular growing interest in the scientific community to understand the principles of brain computation and use them for enabling new treatments for brain diseases  and advance today’s technology with new computational paradigms~\cite{HBP}. Into this direction computational neuroscience studies the information processing properties of the nervous system, developing advanced neural models that require demanding simulation for studying their dynamics.  These simulations are performed usually on High Performance Computing (HPC) clusters~\cite{HPC}, but often slower than the real-time processing of the brain, consuming much more power than any other approach~\cite{HBP}. Even clusters of computers need many minutes to model seconds of worthy brain activity~\cite{Riken_simulation}, given the large amount of studied parameters and drawbacks of the von-Neumann computing architecture~\cite{Indiveri_15}. Neuromorphic systems represent an attractive alternative to conventional numerical simulations, by offering high-speed execution of neural models in silicon with power efficiency~\cite{HBP}. 

The size of neuromorphic systems is steadily increased in order to handle progressively more advanced computational tasks~\cite{Spinnaker, schemmel10, serrano09, davies2018loihi}.
This significantly boosts the requirements for the communication of pulse packets (representing the neurons` action potentials called ``spikes'') on and off the neuromorphic chips, calling for a dedicated communication design capable to provide high bandwidth and sophisticated functionalities.
In recent years several projects developed such large-scale neuromorphic systems, either by using mixed-signal~\cite{schemmel10, Neurogrid} or pure digital approach~\cite{Spinnaker, davies2018loihi}. The requirements of the communication are increased further in the case of accelerated neuromorphic systems~\cite{schemmel10}.
While establishing a high-throughput routing of pulse packets on-chip is feasible, the accelerated neural computation puts significant challenges on the inter-chip pulse event routing.
The inter-chip communication is typically performed by routing of timestamped pulse events that carry information about their realized time and source neuron address \cite{Boahen_2000}.

In this paper we present and characterize such a sophisticated communication infrastructure proposed for the BrainScaleS wafer-scale neuromorphic system~\cite{schmitt2017neuromorphic} which operates at an acceleration factor of \num{10000}. The implementation is centered around an FPGA that employs a digital packet-based network~\cite{Thanasoulis12}. The characterized communication flow is capable to provide reliable configuration data and pulse transmission among the nodes of the system. The implementation includes a dedicated large playback and trace storage level~\cite{Thanasoulis12_DEMO}, capable to provide long-term stimulation and tracing of the resulted neural activity.
The design has a capacity of \num{250e6} events for both stimulation and tracing with a time resolution down to \SI{8}{\ns}.

We assess the communication flow over a neuromorphic wafer unit~\cite{schmitt2017neuromorphic} in terms of throughput, transmission delay, pulse loss and jitter. Our verification methodology goes one step further by evaluating the impact of pulse communication-driven distortions on neural network experiments.
The overall characterization methodology provides an important guideline for the mapping process of the implemented neural model onto the system before an experiment~\cite{bruederle11,muller2022operating}. The communication
characterization allows the minimization of the distortions of the initial stimulus spike trains, ensuring the successful execution of an experiment. The results show that the proposed design meets the needs of spike-based learning experiments, which require accurate, long-term stimulation with iterative fashion~\cite{Thanasoulis_14}.
Our implementation constitutes the most versatile synchronous, serial packet-based communication of timestamped events proposed so far, with full communication characterization and demonstration of an applied neural benchmark.

Section~\ref{sec_material} provides an overview over the employed hardware material and derives the communication requirements.
In Section~\ref{sec_methods} the applied methodology is presented.
Section~\ref{sec_characterisation} includes the characterization of the communication flow.
The implications of the evaluation process are tested with an application of a demanding neural benchmark in Section~\ref{sec_benchmark}.
Finally, some issues worthy for discussion are raised in Section~\ref{sec_results}.

%% file: 02_Material_and_Problem_Statement.tex
	
\subsection{Wafer-scale Neuromorphic System}
\label{wafer_system}

The \textit{BrainScaleS} wafer-scale system developed in the EU projects FACETS \cite{FACETS_EU_FP6_15879}, BrainScaleS \cite{BRAINSCALES_EU_FP7_269921} and the Human Brain Project (HBP) \cite{HBP_web} aims to provide a neuromorphic computing platform to accelerate computational neuroscience and to study its use of energy-efficient computing and AI.
A module of the wafer-scale system is depicted in Fig.~\ref{fig_system}. It employs wafer-scale integration technology to gain a high connection and component density.  The wafer unit of \SI{20}{\cm} diameter has been designed for integrating a maximum amount of neurons and synapses at sufficient flexibility.  
It is composed by 384 neuromorphic chips called High Input Count Analog Neural
Networks (HICANNs)~\cite{schemmel10}, which are mixed-signal
application-specific integrated circuits (ASICs).
They include in total up to $2\times10^5$  silicon neurons implementing the Adaptive Exponential Integrate\&Fire neuron model~\cite{MILL10} and $44\times10^7$ learning synapses.
The system is designed for operating at a nominal acceleration factor of $10^4$ compared to biological real time. This fact increases the simulation speed and integration density of the analog neuron and synapse circuits at the same time. 

The connectivity and model parametrization is achieved by a two-layer approach. 
On each wafer module, individual synapse-and-neuron blocks, mentioned before as HICANNs, are connected directly via a high density routing grid~\cite{schemmel10}, the so-called \textit{Layer-1} network.  Pulse communication among the single dies of a wafer is achieved via this asynchronous and serial event network, which is directly implemented on the wafer and it realizes the intra-wafer communication. 
A dedicated synchronous packet-based network implements the off-wafer communication, connecting the wafer to the surrounding system. External configuration and pulse stimulation as well as monitoring and control of the system is carried out through this high-speed communication network, called~\textit{Layer-2}, on which we focus in this work. This network is centered around an field-programmable gate array (FPGA) environment, allowing for high flexibility and rapid prototyping. The wafer-scale system composes a unique  energy-efficient research platform to study the dynamics of large-scale biologically inspired neural networks with an equivalent resolution in order 
of seconds over a period of many hours in biological real-time. This way standalone simulations of neural networks can be accelerated immensely.

\begin{figure}
\centering
\includegraphics[width=3.3 in]{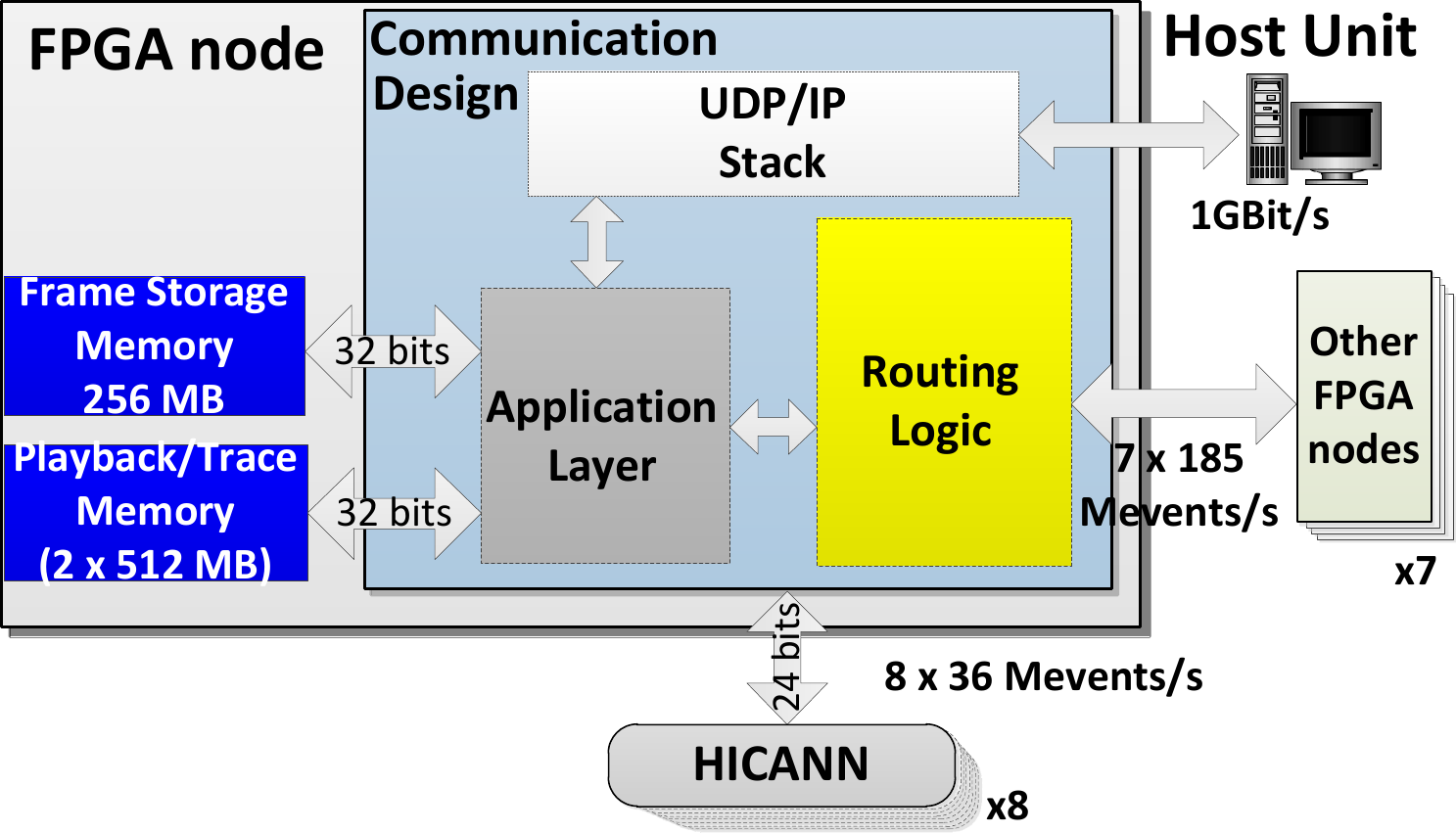}
\caption{The schematic of the Layer-2 network with an FPGA Kintex-7 is interfaced with
eight HICANNs on the wafer and seven surrounding FPGA nodes.
}
\label{fig_layer2}
\end{figure}

\subsection{Layer-2 Network}
\label{layer2}

The Layer-2 network implements the off-wafer communication composed of 48 FPGA nodes, each depicted 
in Fig.~\ref{fig_layer2}, for every wafer module. It realizes the communication with the host unit, the configuration of the system and the pulse transmission among the wafer modules. Each node realized with a Xilinx KINTEX-7 
FPGA~\cite{kintex7} is interfaced with eight HICANNs and (optionally) seven surrounding FPGA counterparts. 
This tree-like structure enables one FPGA to control up to eight HICANNs on the wafer that correspond 
to 4096 silicon neurons. The communication is realized with data packets that contain the essential 
information~\cite{Thanasoulis12}. Data from the host unit or other nodes are forwarded via the FPGA to eight HICANNs and pulse events are routed backwards from the HICANNs following the same route.

Each Layer-2 node provides, as Fig.~\ref{fig_layer2} shows, 1~Gbit/s \textit{Ethernet} adapter for its 
connection with the \textit{host unit} and $7 \times 6.25$~Gbit/s for connections to its counterparts; 
three connections concern local nodes realized direct via the PCB and four connections are dedicated to 
distant nodes implemented with USB 3.0 connectors.
Each Layer-2 node houses two \textit{DDR3} \textit{SDRAMs} of different sizes: a $256$~MB frame storage/ routing memory and 2 $\times$ 0.5~GB \textit{SODIMM} for storing playback and recording data respectively.

The communication with the wafer (8 HICANNs per FPGA) is realized via 8 \texttt{FPGA-HICANN} channels implemented by a \textit{Low Voltage Differential Signaling (LVDS)} transmission of one data and a clock line per direction.
Each LVDS channel runs with \SI{500}{\mega\hertz} in Double Data Rate (DDR) mode, allowing a raw data rate of 1~Gbit/s.
The FPGA-HICANN channels can handle both configuration data and pulse events in both directions.
Data is transmitted in chunks of 8 bit.
Multiple chunks are combined to packets containing an 8~bit header describing the packet type, a variable size payload depending on the packet type, and a 8~bit footer with Cyclic Redundancy Check (CRC).
There is an idle time of two chunks between the transmission of two subsequent packets.
Pulse events consisting of 15~bit timestamp and a 9~bit neuron label require 3 chunks and can be transmitted either as single or double pulse packets consisting of 5 respectively 8 chunks in total. Given the 2 two-chunk idle time, the transmission is limited to 17.86~M~events/s for single and 25 M~event/s for double pulse packets.
The downstream transmission from FPGA to the HICANN uses only single pulse packets in this work, while the upstream transmission uses both types.

In the HICANN the pulse events are processed as follows:
The \emph{15~bit timestamp} refers to a \SI{250}{\mega\hertz} clock running on the HICANN, while the \emph{9~bit label} refers to a 6-bit neuron id and a 3-bit channel id. If \emph{timestamp mode} is enabled, the pulses are only released when the timestamp matches the HICANN clock. Otherwise the pulses are processed immediately.
The 3-bit channel is used to transmit the source neuron id on one of 8 so-called $Layer~1$ buses, which according to a fixed routing forwards the pulse to one or more neuro-synaptic cores targeting many synapses \cite{fieres2008}.
In turn, when one of the 512 physical neurons on the HICANN spikes, a pulse event with a 6-bit neuron id is generated and a 3-bit channel id is added after processing through a merger tree that combines spikes from different neuron blocks. Before transmission to the FPGA, a 15-bit record timestamp is added.
The HICANN supports a \emph{loopback mode}, where external pulse events are looped back to the FPGA receiving a new timestamp but maintaining the same neuron id.

The earlier version of the Layer-2 network for the BrainScaleS wafer included only twelve off-chip nodes, each realized by a Pulse Communication Subgroup (PCS)~\cite{HART10} composed of 4 full-custom ASICs, called Digital Network Chips 
(DNCs)~\cite{ScholzeDNC_12} and an FPGA Virtex-5. Each DNC controlled the transmission to/from 8 HICANNs, resulting to 32 controlled HICANNs for every PCS.

\subsection{FPGA Communication Design}

A dedicated FPGA design~\cite{Thanasoulis12} implements a digital packet-based network that realizes both the host communication and the inter-wafer connectivity.
The employed UDP network stack is advanced with a custom-designed Application Layer (AL) that introduces reliability, flow and error control and implements the formation, routing and handling of network data~\cite{Thanasoulis12}. It includes also dedicated controllers and interfaces to memories and other FPGA nodes. The architecture demonstrates two independent datastreams to/from the neuromorphic wafer, allowing for high throughput, negligible data collision and flexibility in verification. A variety of packet formats are liable for providing 
a variety of services to fulfill the system requirements for pulse stimulation and routing. The transmission of digital packets enables the design to embed the configuration data for the neuromorphic wafer in the regular spike data
stream. This is a significant advantage and is  in contrast with other proposed solutions, which usually employ separate configuration interfaces~\cite{serrano09},~\cite{berge07},~\cite{fasnacht11}.
As mentioned above, the transmitted pulse represents the source neuron address and a timestamp.
This allows the system to configure individual transmission delays, emulating long-range brain connectivity. 
The feature of propagation delays is
very important for various kinds of neuromorphic computation~\cite{Meyer08},~\cite{deco09} and influences
highly the information processing and dynamics of coupled neural groups~\cite{Dahlem09}. A sophisticated routing 
logic at the kernel of the architecture ensures multicast pulse transmission and the implementation of 
transmission delays towards to target neurons of the system~\cite{Thanasoulis_14b}.
However a detailed description of these routing mechanisms is beyond the scope of this paper that focuses on the communication characterization.

\begin{table*}[ht!]
\centering
\footnotesize
\begin{tabular}{|c|c|c|c|c|}
\hline

	\textbf{Memory} & \textbf{Memory}   &  \textbf{Storage} &  \textbf{Write} &  \textbf{Read}      \\

           \textbf{Type} &  \textbf{Size}   & \textbf{Capacity} & \textbf{Bandwidth}$^1$ & \textbf{Bandwidth}$^1$        \\
\hline
	      SDRAM DDR3 & 1GB &  $2\times125 Mpulses$   & 1.6 GB/s  &  1.3 GB/s    \\
\hline 
\hline
	  \textbf{Average} &  \textbf{Stimulation}    & \textbf{Stimulation} &  \textbf{Tracing}  &  \textbf{Pulse} \\

           \textbf{Stimulation} & \textbf{Precision}    & \textbf{Throughput} & \textbf{Throughput}  &  \textbf{Size}\\
    \hline
	 30.5k pulses/source neuron & 8~ns$^2$   & 121 Mevents/s &   250 Mevents/s  & 32 bits\\
		
 	\hline
\end{tabular}
\caption{ Characteristics of the implemented Playback and Trace storage level. \newline
$^1$ it concerns the maximum read and write supported by the physical memory \newline 
$^2$  given in technical domain (0.08~ms translated in biological domain)}
\label{tab_Playb_Trace}
\end{table*}

\subsection{Playback and Trace Storage Level}
\label{sec_playback_trace}

One important feature of the proposed communication design is a large Playback/Trace storage level, which other 
similar solutions~\cite{serrano09},~\cite{fasnacht08},~\cite{berge07} have not implemented.
This storage level is very important for the accelerated neuromorphic computation, 
as the Ethernet bandwidth between FPGA and host computer (Fig.~\ref{fig_layer2}) is too low to support the required data rates for stimulus and spike recording.
For this purpose 2$\times$ 512~MByte DDR3 SDRAM physical separated memories on board are 
employed; the first as \textit{Playback} memory for stimuli pulses and the second as \textit{Trace} memory for 
recording the produced neural activity during an experiment. This way pulse stimulation and tracing are performed
without the mediation of the slower host control, allowing for long-term and standalone operation required 
for instance on learning experiments~\cite{Thorpe_08}. The Playback and Trace storage level allow for 125 Mpulses 
for stimulation (playback mode) and another 125 Mpulses for the tracing mode, since the size of the memories is 
utilized from each mode with 32 bits per pulse.  This load of pulses is responsible for the stimulation from up to 4096 source neurons and 
tracing of 4096 silicon neurons (eight HICANNs) of a wafer unit. With available pulse stimuli of this mount we can 
stimulate 1000 neurons at 10 Hz for a duration of 1.3 seconds execution time in the accelerated system that 
corresponds to 3.5 hours in biological real time.\footnote{10 Hz rate stimuli correspond to 600 pulses/min or 126k pulses at 3.5 h for a neuron}

\begin{figure}
\centering
\includegraphics[width=0.9\columnwidth]{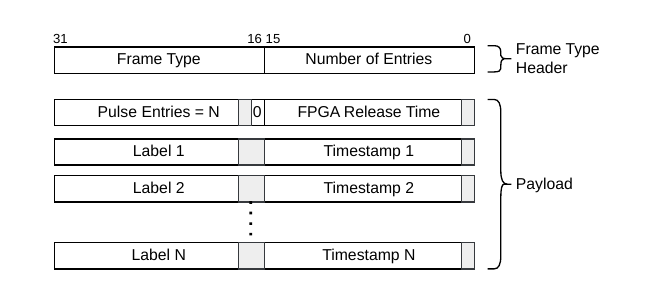}
\caption{The used frame format of the application layer (AL) for the playback memory. In this example the payload is one pulse group with a shared \texttt{FPGA release time} and N pulse with a 14-bit \texttt{label} and 15-bit \texttt{timestamp} considered in the HICANN.
}
\label{playb_frame}
\end{figure}

The stimuli pulses for an implemented spiking neural network are stored initially into the Playback memory 
before the experiment according to a software description at the host unit. Fig.~\ref{playb_frame} shows 
the format of the stored frames at the Playback memory. Each frame consists of the frame type header (32 bit) 
and the payload of pulses composed into different \emph{pulse groups}. Each pulse is released 
according to the so-called FPGA release time included at the beginning of each group. 
When the host unit triggers the experiment start, the stimuli pulses are forwarded to 
the HICANNs as input events for the neural network. The attached timestamp of every group is constantly 
compared with a reference system counter and the pulses in a pulse group are released, if the timestamps are equal or the 
difference is smaller than a pre-configured limit. 
The \emph{blockread} module implements pre-fetching of frame data from the Playback memory in order to avoid wait times due to memory requests.
This way the stimulation is precise with accurate pulse release towards to the HICANNs.
Consecutive pulses with different release time can be forwarded from the Playback memory every six clocks 
cycles that corresponds to 48 ns in technical time or 0.48 ms in biological time domain, due to necessary 
processing steps (access of blockread module, timestamp comparison, target decoding)~\cite{Thanasoulis_14}. 
The temporal resolution of pulse stimulation in the FPGA is 8~ns (equivalent to 0.08~ms in biological time domain)
defined by the FPGA clock frequency (125 MHz), because pulses of each group are forwarded from the 
Playback module in every clock cycle.

The Trace memory records constantly all the upstream pulse activity 
during an ongoing experiment. An overflow packet is stored at the Trace memory every time
when the reference counter is reset, in order the procedure to set correct tracing time for the constantly received pulses. The traced data is read-out afterwards and forwarded at maximum frame size for optimized throughput to the host unit for further analysis of network behavior.

A \textit{playback loop-mode} of the stored stimulation is also supported. It sends all the pre-stored stimuli pulses sequentially again to the HICANNs, resulting in a continuant and periodic stimulation. 

The interface to the physical memory from the FPGA side is optimized to support high-clock rates and efficient bandwidth. It includes a Xilinx Native Port Interface (NPI)~\cite{xilinx_NPI} 
component for supporting burst transfers and dedicated controllers that in conjunction with FIFOs are pre-fetching the stored data in order to reduce the time access to the memory. 
Considering that the number of pulses per pulse group is limited to 184 in our test software for historical reasons,
and a pulse can be sent every clock cycle with an initial overhead of 6 clocks, the stimulation 
throughput is estimated at 121 Mevents/s (184/(184+6)$\times$8ns). 
The Trace module allows for maximum throughput up to 250 Mevents/s by recording two pulses per clock.
Table~\ref{tab_Playb_Trace} summarizes the most important features of the Playback and Trace 
storage level.

\subsection{Running an Experiment} 

Each neuromorphic experiment on the BrainScaleS wafer unit comprises three phases:
\begin{enumerate}
    \item \emph{Hardware Preparation:} The parameters of neuron-synapse-blocks are configured in order to implement the desired spiking neural network model, and all the stimulus pulses are stored to the Playback memory from the host. Optionally, for multi-wafer experiments, the routing tables of the packet-based network are set.
    \item \emph{Experiment:} All FPGA nodes and the clock counters of the involved HICANNs on the wafer are started synchronously. In the FPGAs, this will start the process of releasing stimulus pulses from the Playback memory, as well as the recording of traced pulses originating from hardware neurons on the wafer into the Trace memory. If activated, pulses will be routed between FPGA nodes. After a pre-defined duration, the playback and tracing of spikes is disabled.
    \item \emph{Post-processing:} The traced pulses and optionally changed parameters of the neuron-and-synapse blocks are read back to the host computer. The traced pulses are post-processed to obtain spike times in biological time sorted per neuron allowing for further spike train analysis.
\end{enumerate}

For further details about the experiment execution and the involved high-level software for defining and mapping spiking neural network models to the BrainScaleS hardware, see \cite{muller2022operating}. We further note that the emulation of the spiking neural network on the wafer takes place no matter if the Playback and Trace modules are enabled. After configuration of the HICANNs, the neurons on the chip may communicate with each other, e.g., based on internal reciprocal stimulation. This was shown in \cite{schmidt2025demonstrating} where maintained spiking activity was measured on the wafer even one hour of wall time after sending the initial stimulus spikes / the actual experiment.

\subsection{Neural Communication Requirements}

Requirements on communication bandwidth increase significantly when moving to a large-scale hardware system 
like~\cite{schemmel10}. The high integration density of the system and the accelerated computation
of all neuron and synapse circuits  by a factor of $10^4$ compared to biological time,
result to a pulse activity of particularly high frequency more than any other real time similar processing system~\cite{Spinnaker, serrano09}.  The high activity rate imposes restrictions on bandwidth that should be investigated and taken thoughtfully into account. Very large networks need to be well-structured in order to not exceed the bandwidth for inter-wafer communication~\cite{Partzsch_12},~\cite{Partzsch_11}. The bandwidth of the off-wafer communication could be optimally utilized  when the incoming connections from outside are well-distributed over the wafer and the implemented  networks show also some kind of localized connectivity. The characterization of the communication gives insights for the limits and strengths of the communication. This is an important guideline for advanced neuron placement and routing algorithms that map the 
neural network on the system before the experiment~\cite{bruederle11}.

However, a general estimation of the pulse rates beforehand is impossible to be completed, because this depends on the implemented network and the parameters of the emulated  neural model itself. The communication handles timestamped pulse events that are described by a statistical distribution, which influences highly the network behavior and must be kept as much as possible intact from numerous communications distortions. A running experiment must ensure no or tolerable pulse loss in the transmission that may occur by a high spike count together with a low temporal spread~\cite{Thanasoulis_14}. 
This is important because otherwise the firing rates or the temporal characteristics of a spike train could be easily 
violated with negative impact on a performed experiment like~\cite{Thorpe_08}.

\subsection{Communication Distortions}
\label{distortions}
The transmission of pulses in the presented communication design can affect the neuromorphic experiments in various ways. We assess 3 different distortion types that impact the interpretability of the recorded spikes of spiking neural network emulation.
\begin{itemize}
    \item \emph{Stimulus distortion:} The playback and downstream transmission may distort the input spike trains seen by the neural network on the wafer. Input spikes are typically defined in biological time with a resolution of \SI{0.1}{\ms}. The spike times are translated into the accelerated domain ($10000\times$ smaller, i.e., \SI{10}{\ns} resolution) and packed into pulse groups of the playback memory.  Here, spikes from up to 4096 spike sources for 8 HICANNs are combined into pulse groups with shared FPGA release time (\SI{8}{\ns}). Considering the minimum time of \SI{48}{\ns} between pulse groups, this can lead to distortions of release times in the FPGA. Note that the downstream transmission delay can be considered for the calculation of the FPGA release time such that single pulses reach the HICANNs as requested. Yet, in the downstream transmission there may be pulse loss or additional pulse delay (leading to jitter) if FPGA-HICANN channels are overloaded. All this combined leads to the neurons on the wafer seeing a different input stimuli than defined by the user.
    \item \emph{Distorted spiking neural network output:} Based on the changed (distorted) input stimulus, the behavior of the neural network running on the neuromorphic wafer might change, leading to different output spikes compared to the case when receiving a non-distorted stimulus.
    \item \emph{Tracing distortion:} The tracing, i.e., the upstream transmission of pulses and their recording into DRAM, may distort the output spikes of the neurons. First, in the HICANN, where the 15-bit timestamp are assigned, some spikes might either be dropped  or receive a delayed timestamp due to bandwidth limitations. Then, in the FPGA, some pulses may be dropped due to limited buffer sizes \cite{thanasoulis2019phd} if there are too many pulses from specific HICANNs. Hence, the spike trains recorded in the trace memory may differ from the spike trains on the wafer.
\end{itemize}

A performed hardware experiment could be influenced from all the above distortions, which severely limits the validity of any data analysis applied afterwards.
Obviously, the goal is to eliminate or at least minimize the \emph{stimulus} and \emph{tracing} distortions such that the traced spikes match the spikes occurred on the wafer.
However, in many cases it will not be possible to completely get rid of these distortions, hence, a careful characterization of the communication is needed, 
in order to quantify the different distortion effects.
A verification methodology is presented in the following section.

%% file: 03_Methods.tex
\subsection{Methodology for Tests}
\label{subsec_method_tests}

\begin{figure}
\centering
\includegraphics[width=2.9 in]{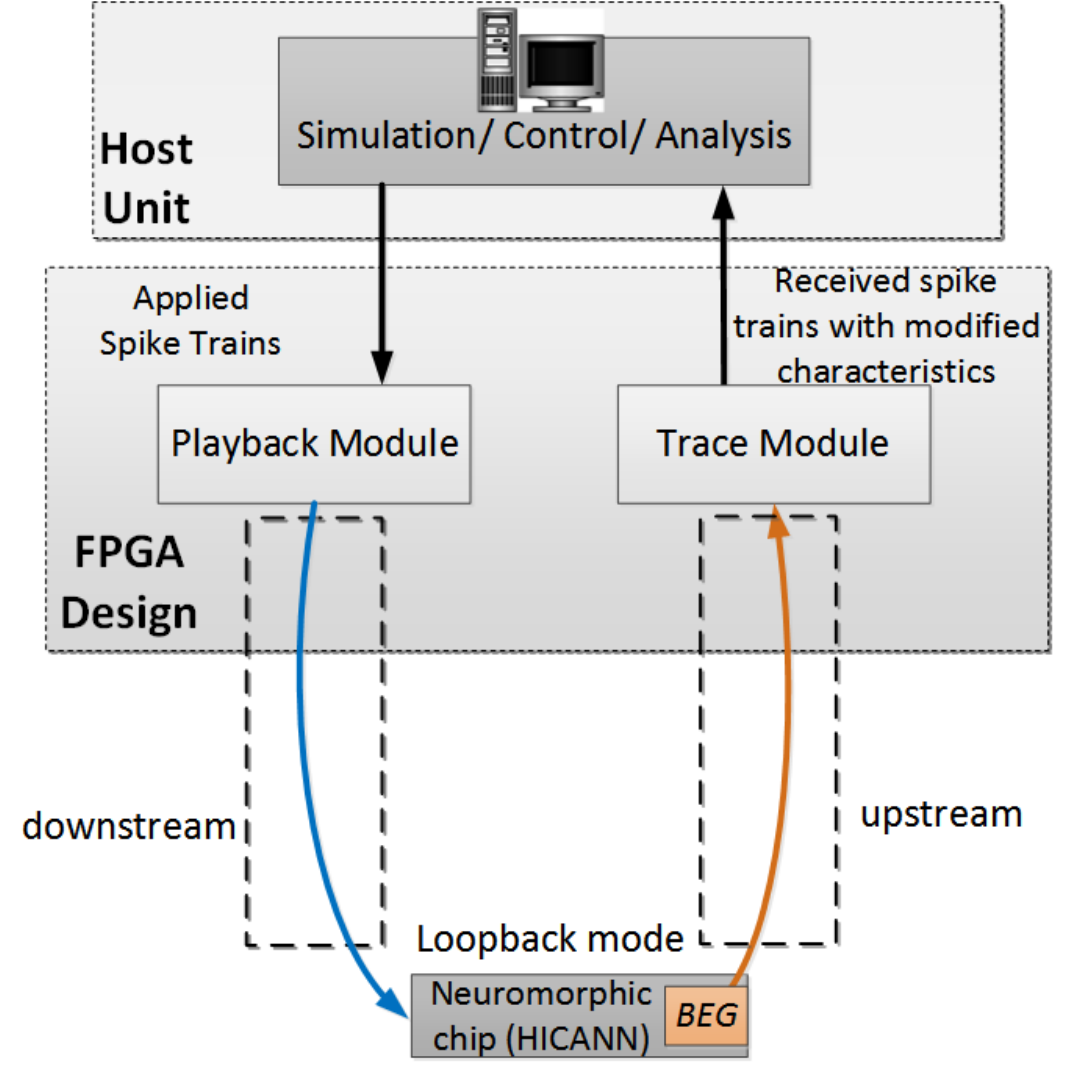}
\caption{The verification scheme for the evaluation of the BrainScaleS off-wafer routing. 
\emph{Characterization of the upstream:}
The HICANNs generate either regular or pseudo-random spike trains which are transmitted to the FPGA and recorded to the Trace memory.
\emph{Characterization of the downstream:}
Pre-generated input pulses with release timestamps are first stored in the Playback memory from where they are transmitted to the wafer.
In the HICANNs they are looped back with a new timestamp to be recorded by the Trace module.
Afterwards, the received and applied spike trains are analyzed on the host to assess the effects of the off-wafer communication.
} %
\label{fig_verif_concept}
\end{figure}

The concept for the verification process of the proposed communication scheme is depicted in Fig.~\ref{fig_verif_concept}.
We focus first on the flow from the HICANNs via the FPGA's Trace module to the Trace memory (\textit{upstream}). 
Then we perform tests on the communication flow between Playback memory, the playback module and the HICANNs (\textit{downstream}).
The chosen order of these tests first on the upstream and then on the downstream is explained because the flow from 
the HICANNs to the Trace module (cf. Fig.\ref{fig_verif_concept}) intervenes also at the tests of downstream for the resulted spike trains that have to be recorded by the Trace module. The verification process is monitored by the host unit.  
Recorded spike trains are analyzed at the host unit, showing any distortions 
on the pulse rate and temporal spread.  
All the measurements for the communication in section~\ref{sec_characterisation} and the application of benchmarks 
in section~\ref{sec_benchmark} have been performed on a neuromorphic wafer module of the BrainScaleS project,
as it is depicted in Fig.~\ref{fig_system}.

For the characterization of the upstream communication spike trains are generated by 
background event generators (BEGs) located on the neuron-and-synapse blocks.
The BEGs mainly have a technical purpose for the operation of the on-wafer
routing network, but can also be used for the stimulation of neurons with
regular or pseudo-random spike trains.
In our tests, the sequence of the pulse events generated by the BEGs is sent to the FPGA
where it is further processed by the Trace module. 
The advantage of the these tests is that the BEGs allow the generation of high spike rates for testing the upstream communication,
which are not bounded by the process of any FIFO or control module like the Playback one. 
This way the limits of the HICANN-FPGA link and the Trace module to record the outcome activity from the neuromorphic 
chips can be revealed.

The characterization of the downstream flow is performed with the so-called~\textit{loopback} mode in the HICANNs. 
Pulses are sent from the Playback memory, via the FPGA design to 
the HICANN chips and then back again to the Trace module (cf. Fig.~\ref{fig_verif_concept}). 
Initially, spike trains with specific temporal characteristics are generated on the host computer either by the simulation of 
a neural model with PyNN~\cite{davison2009pynn} or by custom-written testcases with particular pulse distributions.
Then, these spike trains are loaded to the Playback memory.
After the start of the verification test each pulse is transmitted according to a predefined release timestamp.
When the sent pulses enter a HICANN node they are looped back with a modified timestamp indicating the traced time at the HICANN chip.
Then pulses are forwarded back to the FPGA and recorded at the Trace module.
With the presented approach the off-wafer pulse communication 
can be measured independently of the neural circuits on the wafer. This makes the verification more 
reliable and the implementation simpler. 

The pulses were generated in the biological time domain~(\si{\ms}) and translated to the hardware 
domain~(\si{\ns}) before the measurement and back afterwards assuming an acceleration factor of $\num{10000}$.

The analysis of the data focuses on the pulse throughput, jitter, latency and loss and are explained in the following: 

\begin{itemize}
\item \textit{Throughput}: the data transfer rate that can be carried from the communication design towards
neuromorphic chips in a given time period (usually a second). The throughput is generally expressed in bits
(of data) per second. Additionally, we consider the number of pulses per second by providing the transmitted spike rate in the biological time domain.
\item \textit{Pulse loss}: the fraction of pulses of the original spike train
that is lost on the communication path, so that these pulses either fail to
reach their target neurons, respectively are not traced in the upstream path. 
In our case this may be caused either when we exceed the permitted rate of the network medium (LVDS links) 
or when buffers are full due to bursts occurring in the spike trains. Pulse loss affects negatively
the temporal characteristics and the rate of an applied spike train.
\item \textit{Transmission delay}: the assessment of the delay in the transmission between the releasing point 
(Playback module) and the end node (HICANN chip) is derived by comparing the timestamps of the sent 
and traced pulses.
The initial timestamp corresponds to the desired release time in the Playback module (before packing into pulse groups) and the timestamp of the corresponding traced pulse after the loopback described by the time of the outgoing pulse from the HICANN.
\item \textit{Jitter}: in pulse transmission the jitter is defined as an average of the deviation from the network 
mean latency~\cite{comp_net}. The measurements focus
on the variation (i.e., standard deviation) in the delay. A network with constant transmission delay has no variation and therefore no jitter. 
This variation in transmission delay or pulse arrival times in the receiver (HICANN) characterizes the timing variability. 
\end{itemize}

These characteristics define the quality of service (QoS) of the
proposed communication network and fully characterize  the communication flow.
The tests for the communication include measurements over one HICANN link and eight HICANN links.
This way we can assess the QoS individually for the single connection and for the complete setup.

\subsection{Statistical metrics and spike train analysis}
\label{subsec_spike_analysis}

For the tests several sequences of timestamped pulses are applied, which are called spike trains.
We use either regular or Poisson spike trains:
Regular spike trains have constant spike intervals and are more 
suitable for controlled behavior and measurements for technical 
characteristics. Poissonian spike trains are commonly used for stimulation of neural networks and 
characterize the spontaneous brain activity in the cortex area~\cite{Boustani_10}.
The statistical distribution of the interspike intervals (ISIs) reveal that the neurons discharged  
randomly and independently, explained by uncorrelated Poisson processes~\cite{Bedard_06}.
The  neuromorphic  facility  of the  BrainScaleS project  should be able to support such activity that  enables  the  
emulation of  related theoretical neural models. Therefore tests with Poisson spike trains are
very important for the estimation of the system's behavior and performance.

For the evaluation of the results we use principles and techniques from spike train analysis~\cite{spike_train_analysis},~\cite{Dayan_05}. These are applied specially for the benchmark applications in Sec.~\ref{sec_benchmark}. In neurobiology spike train analysis is used 
to reveal a link between patterns in spike trains and some aspect of neural functioning. 
In our case the same techniques are used to evaluate the traced spike trains and explore 
their temporal coding in comparison with reference spike trains of the initial stimulation. 
The results can show whether the proposed communication flow influences the
characteristics of the applied spike train like rate variation, or the pulse pattern that 
constitutes the statistical distribution. The communication flow should not violate
the distribution of an applied spike train and keep its temporal characteristics intact. 
This can be evaluated by the following techniques of spike trains analysis.
\begin{itemize}
\item \textit{Interspike interval (ISI) distributions} 
are extracted from the recorded spike trains as histograms counting the number
of interspike intervals that belong in discrete time bins.
\item The \textit{coefficient of variation} ($\cvisi$) of the interspike intervals is a measure of the regularity of a spike train:
\begin{equation}
	\cvisi = \frac{ \sigma_\textrm{ISI} } {\overline{\textrm{ISI}}} \label{eq_cv_isi} ,
\end{equation}
where $\overline{\textrm{ISI}}$ and $\sigma_\textrm{ISI}$ are the mean and standard deviation of the interspike intervals of the spike train.
The $\cvisi$ is 0 for a regular spike train and converges to 1 for sufficiently long Poisson spike trains.
\item The \textit{network activity} is the instantaneous total firing rate of
	the whole network, i.e. the total number of spikes from all neurons sampled over a certain
	time bin.
\end{itemize}

%% file: 04_Characterisation_of_the_Communication.tex
\subsection{Verification of the upstream over one node}

\begin{figure}
\centering
\subfigure[]{
    \includegraphics{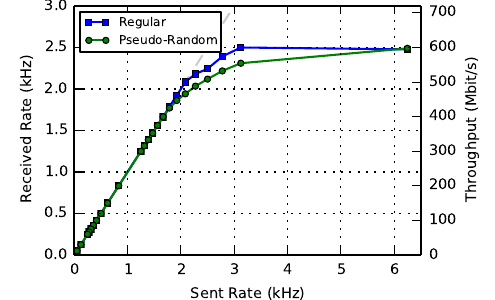}
}
\subfigure[]{
    \includegraphics{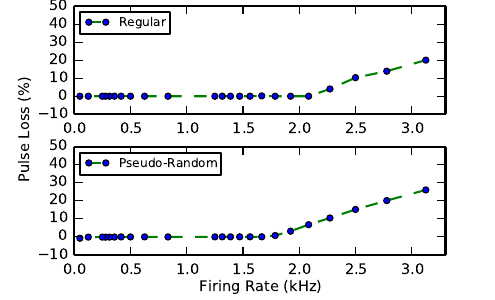}
}
\caption{Pulse throughput (a) and pulse loss (b) with regular and pseudo-random spike trains generated by
background generators on the HICANN chip. The measurement concerns one link for the upstream flow.
The calculation of the throughput assumes 24 bits per pulse.
}

\label{fig_bg_loss}
\end{figure}
The first tests concern the evaluation of the upstream pulse transmission between the HICANNs and the 
Trace module, shown in Fig.~\ref{fig_verif_concept}. 
Spike trains are generated with pre-configured characteristics at one HICANN and are recorded 
on the FPGA. Fig.~\ref{fig_bg_loss} shows the assessment of the throughput and loss in pulse transmission
for regular and pseudo-random spike trains generated by background event generators in a HICANN.
Loss starts to occur at \SI{2.2}{\kHz} for regular spike trains and at \SI{1.8}{\kHz} for pseudo-random 
spike trains. The throughput saturates at 2.5kHz in both cases, corresponding to \SI{600}{\mega bit\per\second} data rate.
This can be explained by the physical limit of the LVDS lines between the HICANN and the FPGA.
There is an idle time of 2 chunks between two packets (cf. subsection \ref{layer2}).
Hence, there can be one single pulse packet every \SI{56}{\ns}, or one double
pulse packet every \SI{80}{\ns}, corresponding to spike rates in the biological
time domain of \SI{1.78}{\kHz} and \SI{2.5}{\kHz} respectively.
Double pulse packets are automatically generated when 2 pulses are buffered for transmission.
The results in Fig.~\ref{fig_bg_loss} prove that this feature works and that the maximum throughput is sustained.

\subsection{Verification of the downstream over one node}

The tests concern the evaluation of the pulse transmission from the Playback module towards 
HICANN chips as the end-node of the verification chain.
Timestamped pulses that correspond to a predefined temporal sequence and
distribution are forwarded by the Playback module to a HICANN node, where the
pulses are looped back to the FPGA and finally recorded by the Trace module,
as was described in Fig.~\ref{fig_verif_concept}.

\subsubsection{Throughput}

\begin{figure}
\centering
\includegraphics[width=2.8 in]{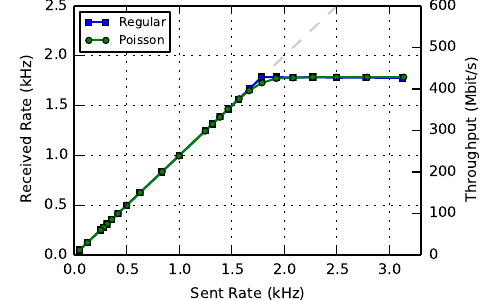}
\caption{Throughput measurements of the downstream pulse flow between the Playback module 
and one HICANN chip using the loopback in the HICANN. The throughput saturates at 430 MBit/s (assuming 24 bits per pulse). 
}
\label{fig_throughput}
\end{figure}

Throughput measurements between the Playback module and the HICANN chip are shown in 
Fig.~\ref{fig_throughput}. It illustrates the rate of received pulses with respect the rate of the stimulation rate.
Taking into account the accelerated computation of factor $10^4$  and 24 bit per pulse the throughput 
(received spike rate) saturates at approximately 430 Mbit/s, due to the limits of the physical transmission links.
This refers to the maximum supplied bandwidth when only single pulse packets are used in the FPGA-HICANN link 
(cf. subsection \ref{layer2}). Note that in our experiment setup, we only use single pulse packets in the 
downstream path, i.e. we prefer timing accuracy over throughput.

\subsubsection{Pulse Loss}

\begin{figure}
\centering
\includegraphics[width=3.3in]{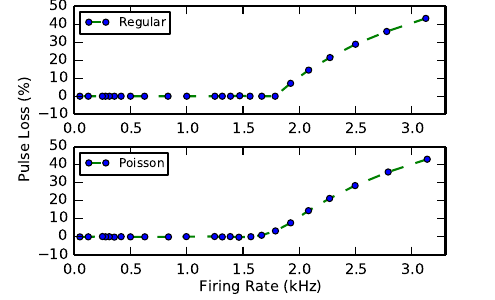}
\caption{Pulse loss of regular and Poisson spike trains in loopback experiments via one HICANN node.
}
\label{fig_loss_figure}
\end{figure}

Spike trains either with constant time-intervals (regular) or generated via a Poisson process were 
applied. The measurements with regard to the pulse loss are shown in Fig.~\ref{fig_loss_figure}. 
The top panel illustrates the results from the tests with regular spike trains. 
From the diagram it is clear that for biological firing rates below 1.8  kHz there is no pulse loss in the 
datastream. When the rate of the applied spike trains exceeds this boundary then the pulse loss is 
increased progressively. The measured loss in this case are caused by the limited bandwidth of the 
physical FPGA-HICANN link.

The results for Poisson spike trains are shown at the lower part of Fig.~\ref{fig_loss_figure}. In this case spike loss starts to occur at $\approx \SI{1.6}{\kHz}$ biological rate. This is explained by the fact that a Poisson process generates random distributions with very small time slices that show high firing rate. 
If the amount of spikes inside these time slices is large this means that too
many spikes had to be delivered to the HICANN chip in a short time interval.
As result many of these spikes have to be buffered at the FPGA side of the FPGA-HICANN link, and this exceeds sometimes the limited buffering capabilities so that pulses are dropped.
The results also show that the loss is increased smoothly in function of the rate for the case of Poisson 
spike trains in contrast to the regular case, where loss starts when exceeding the maximum throughput.
This is explained by the fact that the rate in the Poisson spike trains exceeds the limit at gradually more
time slices.

The results show that there is no pulse data in the data transmission within the implemented FPGA design.
Pulse loss starts to appear only when the applied rate exceeds the boundary rate of the physical 
communication link that is imposed per design. In order to overcome this the pulse communication 
should be distributed properly over several HICANN links.
This ensures that the rate for each link will never exceed the limit.

\subsubsection{Transmission Delay and Jitter}

\begin{figure}
\centering
\includegraphics{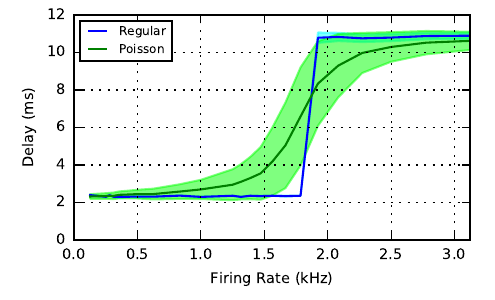}
\caption{Transmission delay and jitter in loopback experiments via one HICANN for sweep over the sent rate: Average delay (solid lines) $\pm$ jitter (shaded area).
}
\label{fig_delay_and_jitter}

\end{figure}
\begin{figure}
\centering
\includegraphics{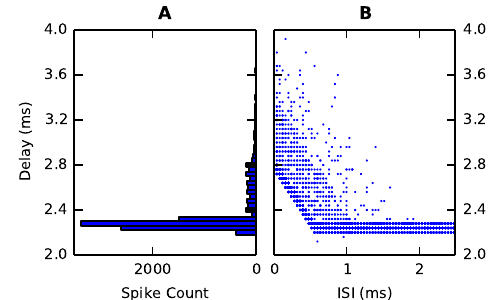}
\caption{Transmission delay and jitter estimation for loopback experiment over one HICANN: (A) Delay histogram and (B) delay as a function of the interspike interval (ISI) with respect to the preceding pulse. Data from a Poisson spike train with 10K pulses and \SI{417}{\Hz} mean rate without any loss. Plot (B) is cropped at \SI{2.5}{ms} ISI.
	Mean delay and jitter: \SI[separate-uncertainty=true]{2.34 \pm 0.20}{\ms}.
}
\label{fig_delay_histogram}
\end{figure}

For the same loopback experiments over one HICANN we analyze the transmission delay,
i.e., the time passed between the nominal release time in the Playback module
and the record time in the HICANN after the loopback, cf. section~\ref{subsec_method_tests}.

Fig.~\ref{fig_delay_and_jitter} shows the average transmission delay and jitter
for the sweep over the sent rate:
For regular spike trains the delay is constant and amounts to \SI{2.3}{\ms}
biological time, corresponding to \SI{230}{\ns} technical time. 
This latency is caused by the time needed for packet handling and transmission via
the physical links of the FPGA-HICANN flow.
The jitter for regular spike trains is negligible (\SI{< 0.04}{\ms}), because
the pulses come maximally \SI{0.12}{\ms} late.
For firing rates higher than the maximum bandwidth of a single HICANN link the
loopback delay suddenly jumps to and saturates at a delay of \SI{10.7}{\ms}.
This can be explained by a buffer size of 16 before the FPGA-HICANN link:
When the buffer is full except for the last entry, it takes another 15 pulses,
hence $15\cdot\SI{56}{\ns} = \SI{840}{\ns}$, until the last accepted spike is
processed. Hence the average delay is $\SI{2.3}{\ms}+\SI{8.4}{\ms} =
\SI{10.7}{\ms}$.

In contrast, for Poisson spikes there is an apparent dependence of the
average transmisson delay and jitter from the sent firing rate
(Fig.~\ref{fig_delay_and_jitter}):
The average delay steadily increases with the firing rate and finally saturates at the
same maximum as for regular spike trains.
This behaviour can be elucidated by looking at the distribution of delays of a single loopback experiment.
Fig.~\ref{fig_delay_histogram}A shows the delay histogram for a Poisson spike train with \SI{417}{\Hz} mean rate, for which no pulse loss occured during transmission.
The peak delay is at \SI{2.3}{\ms}, which is equal to the mean delay for regular spikes.
The delay histogram (A) shows also that a small fraction of pulses has delays up to \SI{2.8}{\ms} or higher.
This effect that resembles a ``tail'' in the histogram can be explained by relating the measured delay of each pulse to the interspike interval (ISI) from the preceding pulse.
The results are depicted in (Fig.~\ref{fig_delay_histogram}B) and we can see that mostly
pulses with an ISI lower than \SI{0.56}{\ms} are affected by this additional
delay, which is caused by the serial processing of pulses in the FPGA-HICANN
link.
There, pulses are processed every \SI{56}{ns}, which results in an additional
delay for those pulses of $(\SI{0.56}{\ms} -\textrm{ISI})$ in biological time.
This effect can accumulate to higher additional delays when there are several
dense pulses in a row.
For the same reason, the jitter increases with the firing rate for Poisson
spikes, as can be seen in Fig.~\ref{fig_delay_and_jitter}:
While for rates below \SI{1}{\kHz} the jitter is smaller than \SI{0.6}{\ms},
for higher rates the jitter even exceeds \SI{2}{\ms}.
This may cause the average delay to be shifted towards \SI{5}{\ms} when a
noticeable amount of pulses is lost at rates larger than \SI{1.6}{\kHz}.
The origin of the saturation of the delay for high rates is the same as for
regular spikes.

The measurement of the transmission delay is very important, because then
stimulus pulses can be released from the FPGA design earlier in order to arrive
at the target neurons on time.
This results in an accurate and reliable stimulation procedure, which is
tolerant to any inherent and inevitable delays of the communication flow. 
The analysis of the jitter provides useful information for choosing the
operation point, e.g., the average bandwidth utilization of the communication path depending on the required QoS.

\subsubsection{Spike train analysis}

\begin{figure}
\centering
\includegraphics{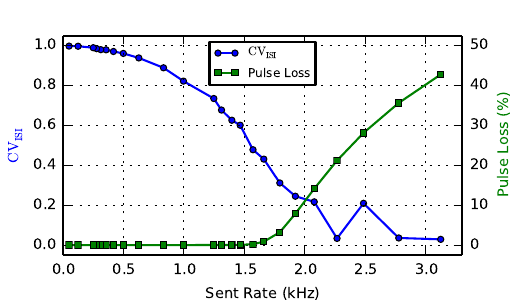}
\caption{The coefficient of variance of the interspike intervals ($\cvisi$)
	of the traced spike trains, with regard to different mean Poisson rates and
	the pulse loss in the case of the loopback experiment using a single
	HICANN.
}
\label{fig_cv_loopback}
\end{figure}

In the following we analyze how the pulse communication path affects the sent
spike trains, therefore we assess the $\cvisi$ (Eq.~\ref{eq_cv_isi}) and ISI histograms of the
original and traced Poisson spike trains.
Fig.~\ref{fig_cv_loopback} shows the impact of the pulse transmission on the
interspike interval distribution. 
The coefficient of variation of interspike intervals is plotted with regard to
the mean rate and the loss.
As the rate increases the $\cvisi$ decreases and strongly deviates from the
original Poissonian characteristics.
Even for rates without any loss, e.g. at \SI{1}{\kHz}, the $\cvisi$ takes
values around \num{0.8}.
When there is significant pulse loss, the traced spike train more and more
turn into a regular spike train, which can be explained by the fact that
pulses are processed successively every \SI{56}{\ns} in the FPGA-HICANN link.

\begin{figure}
\centering
\includegraphics{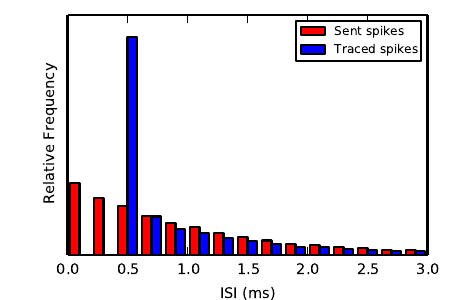}
\caption{Distribution of interspike intervals of a Poisson spike train of
	\SI{1}{\kHz} before and after the loopback experiment over one HICANN.
    The statistics of the traced spike train are different due to the sequential downstream transmission of pulse packets in the FPGA-HICANN channel with minimum technical interval of \SI{56}{\ns} (\SI{56}{\ms} biological time).
}
\label{fig_isi_hist}
\end{figure}

More insights on the hardware specific distortions on the original spike trains
are gained by looking at the ISI distribution of the traced and original
spike train.
Fig.~\ref{fig_isi_hist} shows these distributions for a Poisson spike train of
\SI{1}{\kHz}:
The original spike train has exponentially distributed ISIs, while the
distribution of the traced spike train is clearly distorted:
There are no ISIs below \SI{0.5}{\ms}, instead, they all have been stretched to
the minimal time between two spikes in the FPGA-HICANN link (\SI{0.56}{\ms}).
Another effect is visible for higher ISIs, where the relative frequency of the
ISIs is always lower for the traced than for the original spike train.
This is explained by spike pairs, where the first spike gets an additional
delay while the second spike is processed with the baseline delay, so that the
traced ISI is reduced.

\subsection{Verification of the upstream over multiple nodes}

\begin{figure}
\centering
\subfigure[]{
    \includegraphics{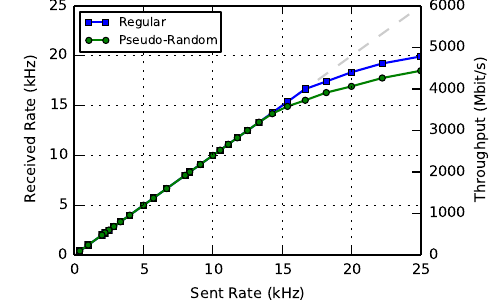}
}
\subfigure[]{
    \includegraphics{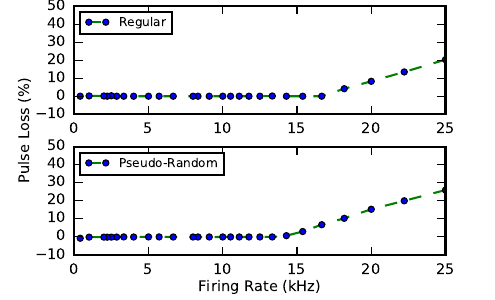}
}
\caption{Verification of the upstream over multiple nodes: Throughput (a) and
	pulse loss (b) for regular and pseudo-random spike trains generated by
	background event generators on 8 HICANNs.
}

\label{fig_bg_loss_8hicanns}
\end{figure}
The evaluation of the proposed communication flow can be extended to 8 links
that connect the FPGA node to 8 HICANN chips respectively as is depicted in
Fig.~\ref{fig_layer2}. 
For the verification of the upstream path we again enable the background event generators (one per HICANN) in regular or pseudo-random mode and we vary the mean
firing rate (same on each HICANN).
The pulse throughput and loss are shown in Fig.~\ref{fig_bg_loss_8hicanns}:
The throughput for regular spike trains saturates at \SI{20}{\kHz}, which is
also approached for pseudo-random spike trains.
This maximum upstream throughput can be explained by the maximum throughput of the
single HICANN connections (\SI{2.5}{\kHz}) multiplied by the number of HICANNs
(eight).
These measurements show that the trace module, which has a theoretical
bandwidth of \SI{25}{\kHz}, can successfully process the spikes from all
HICANNs without any loss.
The pulse loss shown in Fig.~\ref{fig_bg_loss_8hicanns} can be fully attributed
to the processing in the HICANN-FPGA links, considering the findings for one link
in Fig.~\ref{fig_bg_loss}.

\subsection{Verification of the downstream over multiple nodes}

\begin{figure}
\centering
\subfigure[]{
    \includegraphics{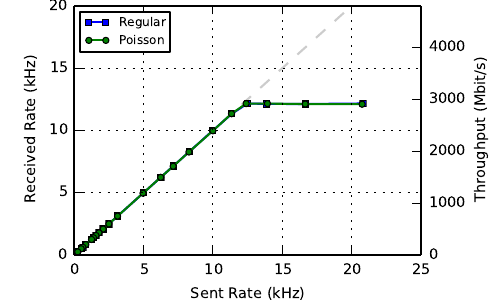}
}
\subfigure[]{
    \includegraphics{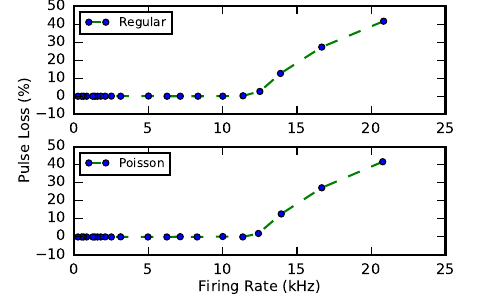}
}
\caption{Verification of the downstream over multiple nodes: Throughput (a) and
	pulse loss (b) for regular and Poisson spike trains in loopback experiments via 8 HICANNs.
}

\label{fig_loopback_loss_8hicanns}
\end{figure}

The downstream communication via 8 HICANN nodes is tested by applying regular
and Poisson spike trains designated to 8 HICANNs via the Playback module in
loopback experiments.
The results are depicted in Fig.~\ref{fig_loopback_loss_8hicanns}:
The throughput saturates at \SI{12.5}{\kHz} for both types of spike trains,
which corresponds to the bandwidth of the playback module (cf. Section
\ref{sec_playback_trace}).
Loss starts to occur equally for regular and Poisson spike trains when this
limit is exceeded. In summary, the downstream communication is mainly limited by the bandwidth of
the Playback module, while the upstream communication is limited only by the
bandwidth of the FPGA-HICANN links, but not by the trace module.

We also evaluated the transmission delay and jitter in the 8 HICANN loopback
experiments (not shown):
The delay and jitter depending on the sent rate over 1 HICANN behave
approximately the same way as in the 1 HICANN experiments.
Only when the sent rate over all HICANNs approaches the downstream limit of
\SI{12.5}{\kHz}, the transmission delay suddenly explodes, as pulses are more
and more shifted by the software that packs the pulses into pulse groups (cf.
Section~\ref{sec_playback_trace} and Fig.~\ref{playb_frame}).

The analysis of the traced Poisson spike trains (1 per HICANN) revealed similar
distortions as in the single HICANN loopback experiments:
Again, there is a minimum ISI of \SI{56}{\ns} and the $\cvisi$ steadily
decreases with the firing rate, as in Fig.~\ref{fig_cv_loopback}.

%% file: 05_Benchmark_Application.tex
The suitability of the proposed communication design for real neuromorphic experiments is assessed with the application of a neural model as benchmark.
We remark that the same model and benchmark methodology has been applied in \cite{thanasoulis2019phd} using the earlier version of the off-wafer network, cf.~Section~\ref{layer2}.

\subsection{Test Approach}
After the systematic characterization of the packet-based Layer-2 network in the previous section, 
we now want to test the pulse communication under realistic conditions.
Especially, we want to test and analyze its performance and suitability when executing neural network 
experiments on the system.
Ideally, this would entail the emulation of a neuromorphic benchmark on the wafer, that is stimulated 
by the Playback module and whose spiking activity is recorded to the trace memory.
We could then study the influence of the Layer-2 on the network behavior, e.g., whether the 
bandwidth limitations in the downstream affect the experiment or how the recorded spiking activity
differs from the activity on the wafer due to potentially pulse loss at the upstream.
However, this would require a reliable operation of the neuromorphic circuits on the HICANNs on the 
one hand, and the complete knowledge about the spiking activity on the wafer on the other hand.
The former could not be guaranteed at the time of performing the experiments (2016), as the calibration of the mixed-signal neurmorphic circuits
was an ongoing and not yet completed process.
The latter is for the design not possible, as the Layer-2 network is the monitor for the spiking 
activity and no alternative is available.

Therefore, a different strategy is needed to test the off-wafer pulse communication architecture with realistic spike trains independently from the neuromorphic circuits on the wafer.
In a previous work \cite{ScholzeDNC_frontiers} a modular feedforward network
was used to characterize the predecessor of the current Layer-2 network (the
pulse communication subgroup consisting of 1 FPGA and 4 DNCs, cf. section~
\ref{layer2}):
the pulse transmission between different neuron groups was implemented in
hardware using a loopback-connector at the DNC replacing the HICANN, while the
neuron dynamics were simulated in software.

Here, we take a different approach: \emph{we simulate a neural network model completely in software, perform a loopback experiment with the recorded spike trains and compare them to the traced hardware pulses.}
The experiment and analysis is repeated for different hardware setups with varying numbers of HICANNs.

\emph{We note the following:} From the characterization in Sec.~\ref{sec_characterisation} we know that the downstream transmission is more limited (due to the playback bandwidth) than the upstream transmission. Hence, running a loopback experiment with spikes from a network simulation will not provide further insights about the pulse tracing limitations -- which would be interesting for real network models implemented on the wafer.
Instead, the benchmark application will mainly measure the effects of the downstream transmission on spatiotemporal spiking activity stemming from real computational neuroscience simulations, instead of investigations of the downstream under simplified conditions with regular or Poisson spike trains.

\begin{figure}[ht]
\centering
\subfigure[Raster plot]{
    \includegraphics[width=0.42\textwidth]{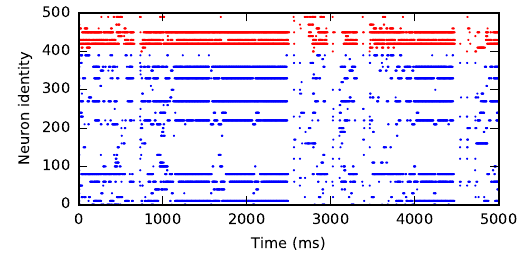}
    \label{fig_raster_plot}
}

\subfigure[Network activity]{
    \label{fig_network_activity}
    \includegraphics[width=0.42\textwidth]{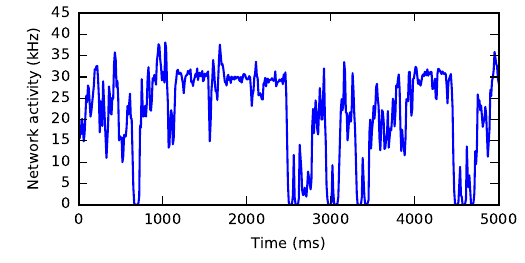}
}

\subfigure[Activity histogram]{
    \label{fig_activity_histogram}
    \includegraphics[width=0.42\textwidth]{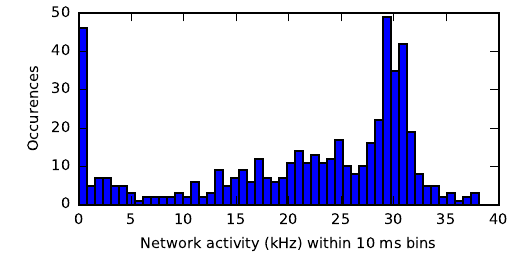}
}

\subfigure[$\cvisi$ histogram]{
    \label{fig_CV_isi_histogram}
    \includegraphics[width=0.42\textwidth]{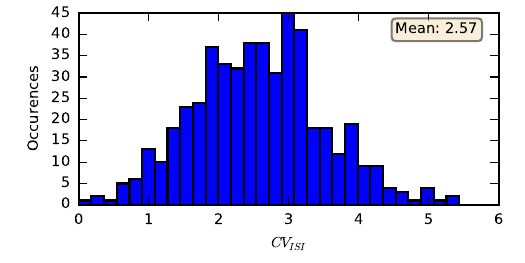}
}

\caption{Simulation and analysis of the benchmark network of 500 AdEx neurons showing self-sustained activity.
(a) Raster plot of spiking activity: excitatory neurons are shown in blue, inhibitory neurons in red, only 10\% of neurons are shown for sake of visibility.
(b) Network activity averaged in bins of \SI{10}{\ms}.
(c) Histogram of the network activity.
(d) Distribution of $\cvisi$ of all neurons.
The network was simulated for \SI{30}{\second}, for plots (a)-(c) only the first \SI{5}{\second} are considered.
} 

\label{fig_AI_simulation}
\end{figure}

\subsection{Benchmark Model}
The chosen network model is a randomly connected cortical network based on \cite{Destexhe09} that shows self-sustained activity switching between asynchronous irregular (AI) and synchronous regular (SR) network firing states.
The network contains 500 conductance-based Adaptive Exponential Integrate and Fire neurons (AdEx, 80\% excitatory and 20\% inhibitory), which are randomly connected with a probability of 8\%. 
The excitatory cells include a proportion of 5 \% low-threshold spiking (LTS) cells, which 
help to maintain self-sustained AI spiking activity for small network sizes \cite{Destexhe09}.
Initially 20\% of the neurons are stimulated for a small time window of 50 ms.
The network was simulated with NEST~\cite{NEST} using the
PyNN~\cite{davison2009pynn} script available in ModelDB \cite{mcdougal2017twenty} at 
\url{https://modeldb.science/126466}.

The spiking activity of the investigated network is shown in Fig.~\ref{fig_AI_simulation} \subref{fig_raster_plot}, exhibiting different network states over time:
At the beginning, the neurons fire in uncorrelated and irregular manner (AI), after 700 ms the network starts to oscillate between states of high AI activity and quasi zero spiking activity (Up and Down states), and enters a synchronous regular state (SR) after 1200 ms with a small fraction of neurons to keep firing with rather constant interspike intervals as  Fig.~\ref{fig_AI_simulation} \subref {fig_raster_plot} depicts. Then the network demonstrates again Up and Down states with oscillations (cf.  Fig.~\ref{fig_AI_simulation} \subref {fig_network_activity}) with the total activity to show little correlation in time. 
The distribution of the $\cvisi$ in \subref {fig_CV_isi_histogram} shows that
most neurons have a $\cvisi$ larger than 1 indicating the switching between
regular and irregular activity.

\subsection{Benchmark Application Strategy}
The above network has a total average firing rate of \SI{19.9}{\kHz} (\SI{39.9}{\Hz} per neuron).
For a network experiment on the hardware, the 500 neurons need to be distributed over different HICANNs.
In loopback mode, up to 256 neurons (spike sources) can be mapped to one HICANN.
It is obvious, that it does not make sense to map all 500 neurons onto just two HICANNs, as the total mean rate of \SI{19.9}{\kHz} is much larger than the maximum throughput 
provided by two HICANN link ($2\times\SI{1.8}{\kHz}=\SI{3.6}{\kHz}$). 
In addition, the maximum bandwidth of the playback module (\SI{12.5}{\kHz}) serving 8 HICANNs must be considered.

The problem is thus to find a distribution of neurons onto HICANNs such that
the spikes are transmitted faithfully over the off-wafer network without wasting
too many resources.
One approach is therefore to use as many HICANNs and FPGAs
as are needed to provide the necessary bandwidth for the whole network.
The question is how many extra bandwidth is needed in addition to the
average network firing rate to guarantee a faithful communication.

In the next section we investigate different mapping scenarios in terms of
number of neurons per HICANN.
E.g., if we match the supplied bandwidth to the average network firing rate, there
cannot be more than 39 neurons per HICANN due to the limit of the playback
module.
Alternatively, we could provide a bandwidth equal to the maximum activity within
a 10 ms bin (cf. Fig.~\ref{fig_AI_simulation} \subref{fig_activity_histogram})
of around \SI{40}{\kHz}, which would require the double of resources than
before.

To examine the full range of mapping scenarios, we map between 5 and 64 
neurons per HICANN, and perform the loopback experiment as described in
section~\ref{subsec_method_tests}.
Note that for the complete network of 500 neurons between 8 and 100 HICANNs are
needed depending on the number of neurons per HICANN.
Loopback experiments involving several FPGAs were not executed simultaneously,
but rather sequentially on the same physical system.
After this the traced spike trains are merged and compared to the original network in
terms of pulse loss, network activity, and irregularity of the spike trains.

\subsection{Benchmark results}
\begin{figure}
\centering
\includegraphics[width=3.5in]{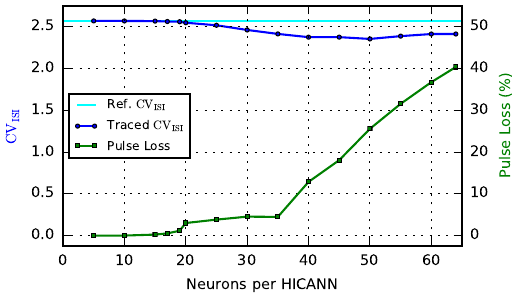}
\caption{Benchmark application: pulse loss for the sweep of different numbers of neurons per HICANN of the AI states model~\cite{Destexhe09}.
}
\label{fig_AI_neurons}
\end{figure}

The results of the benchmark application in a sweep over the number neurons per HICANN are shown in Fig.~\ref{fig_AI_neurons}. 
Pulse loss starts to occur for more than 15 neurons per HICANN but stays below \SI{1}{\%} until it jumps to \SI{3}{\%} at 20 neurons.
Then from 20 to 35 neurons per HICANN the pulse loss increases only slightly from 3 to 4\%.
For more than 35 neurons per HICANN the pulse loss rises more sharply, which can be
attributed to the fact that the supplied bandwidth is below the average firing rate
of the whole network such that pulse loss is inevitable.
The figure also shows the mean $\cvisi$ of the reference and traced spike trains: starting from 20 neurons per HICANN the $\cvisi$ starts to deviate from the reference, going down from $~2.6$ to $~2.4$ when the pulse loss rises. Overall, the $\cvisi$ still shows a high value indicating switches between phases of regular and very irregular activity.

\begin{figure}[ht]
\centering
\subfigure[Results from tests with 15 neurons per HICANN ]{
    \includegraphics[width=0.49\textwidth]{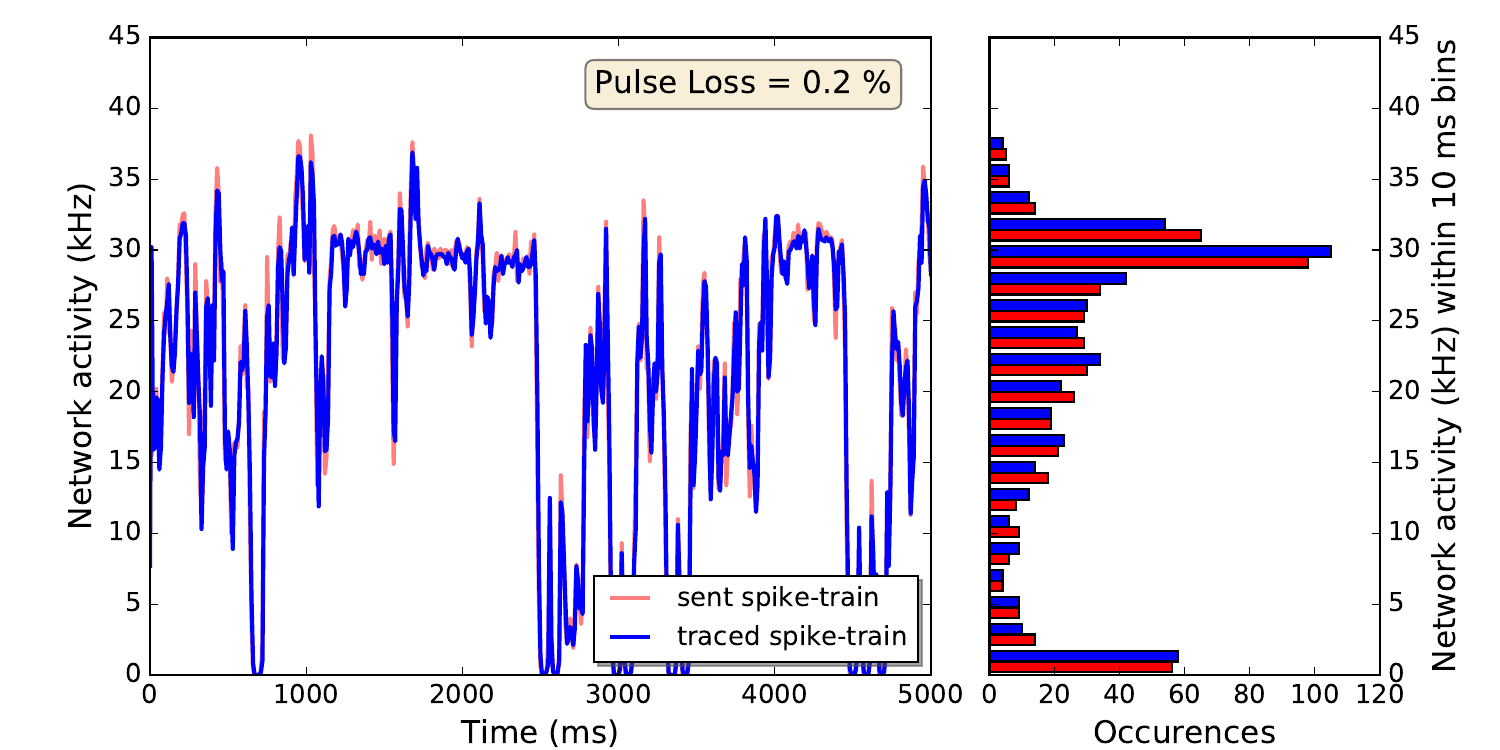}
    \label{fig_ai_15_neurons}
   
}
\subfigure[Results from tests with 20 neurons per HICANN]{
    \label{fig_ai_20_neurons}
    \includegraphics[width=0.49\textwidth]{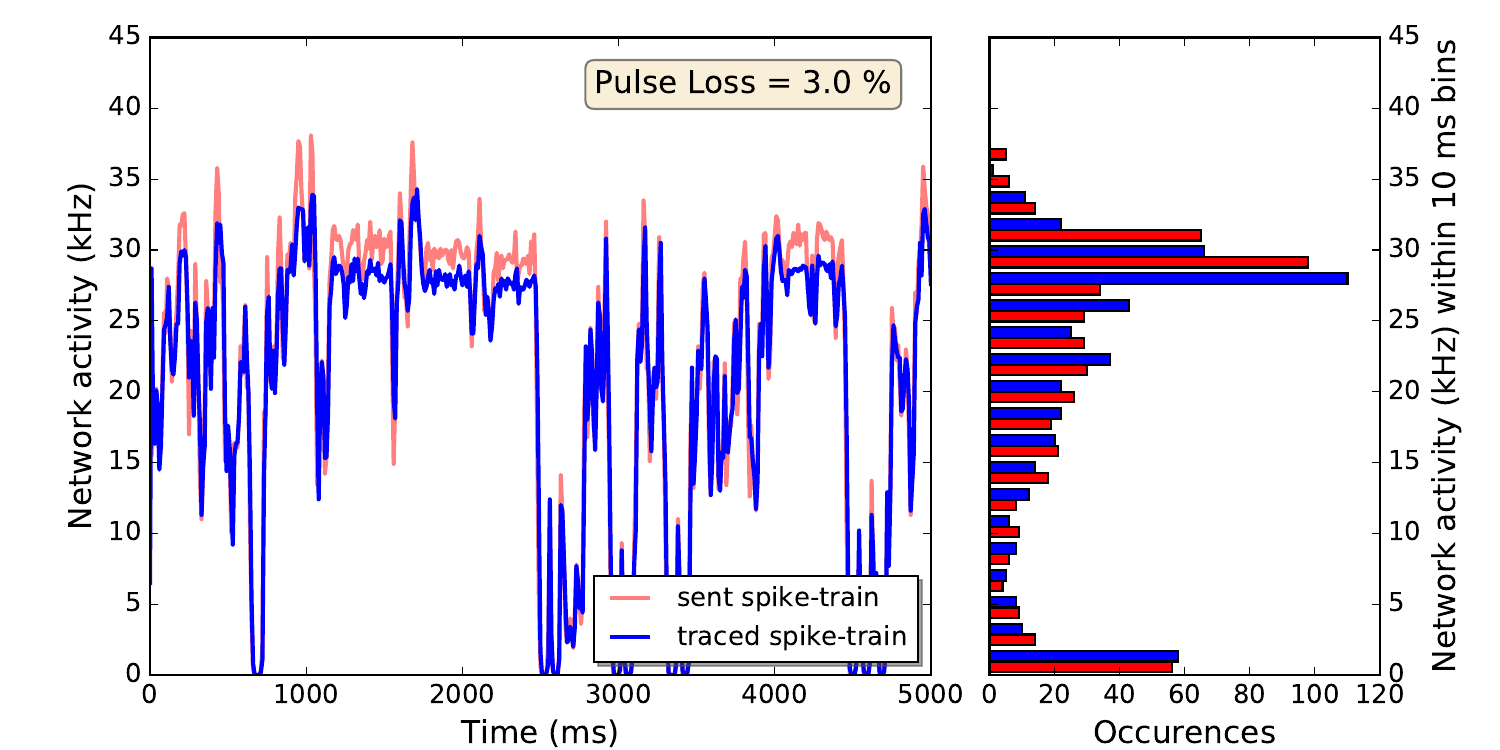}
}
\subfigure[Results from tests with 40 neurons per HICANN]{
    \label{fig_ai_40_neurons}
    \includegraphics[width=0.49\textwidth]{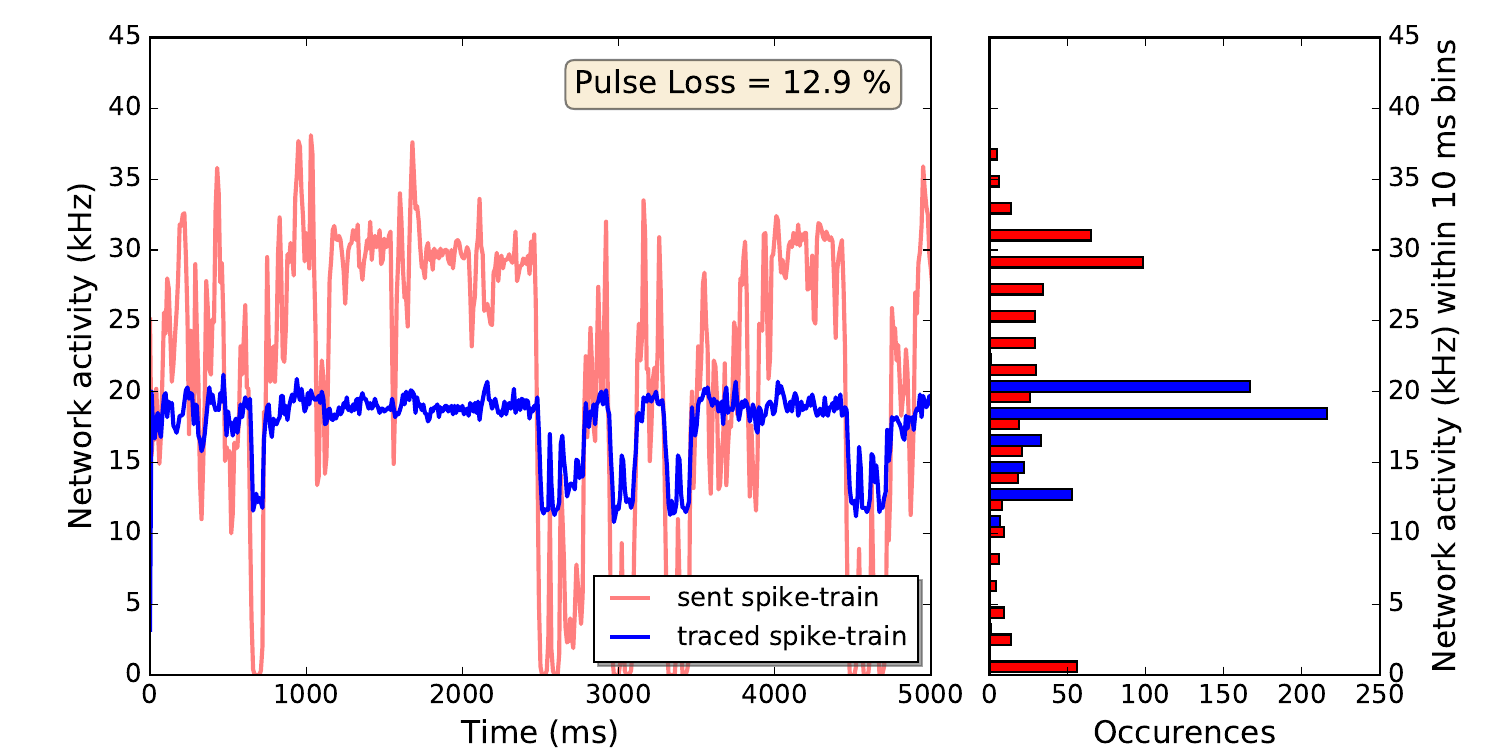}
}
\caption{Analysis of network activity for three applications of the benchmark model with different distributions of the neurons across HICANNs:
	(a) 15, (b) 20 and (c) 40 neurons per HICANN.
\label{fig_AI_analysis}
}
\end{figure}

We study the cases with 15, 20 and 40 neurons per HICANN in more detail:
Fig.~\ref{fig_AI_analysis} shows the reference network activity from Fig.~\ref{fig_AI_simulation} and the same activity after its application on the Layer-2 
communication network.
The \emph{sent spike train} refers to the reference activity and the \emph{traced spike train} to the pulses recorded into the Trace memory in the loopback experiment.
The results for 15 neurons per HICANN are shown in Fig.~\ref{fig_AI_analysis} (a) with the reference network activity nearly matching the recorded activity. 
This can also be seen in the activity histogram in the right panel.
The average rate per HICANN is \SI{\approx 0.6}{\kHz} and the pulse loss is close to 0.
We suspect that the small deviation is caused by additional delays in the transmission of subsequent pulses.

Fig.~\ref{fig_AI_analysis} (b) shows the case of 20 neurons
per HICANN and a pulse loss of \SI{3}{\%}.
Differences mainly occur during phases of high spike rates, but the overall pattern of spike activity is maintained.
Although the \SI{\approx 0.8}{\kHz} mean rate per HICANN node is significantly below the single HICANN bandwidth (\SI{1.8}{\kHz}), pulse loss can occur due to locally and temporarily much higher spike rates.

In Fig.~\ref{fig_AI_analysis} (c) the experiment with 40 neurons per HICANN and \SI{12.9}{\%} pulse loss shows strong deviations between sent and traced spike data.
At no time does the average firing rate of the traced spike trains match the original one.
Overall, this leads to strongly distorted network activity, which is also notable in the histogram on the right.

Consequently, taking into account the performance measurements of 
section~\ref{sec_characterisation} and applying these tests 
we conclude that for a network distribution of 15 or 20 neurons per HICANN the Layer-2
network is suitable to provide reliable communication for the execution of the neural
benchmark based on~\cite{Destexhe09}. 
Note that at these mapping scenarios the average utilization of the bandwidth
of the Playback module and the HICANN link is less than \SI{50}{\%}, which
might present a good point of orientation for other neural network
experiments.

%% file: 06_Discussion.tex
\subsection{Summary and Limitations of Current Design}
In this work we characterized the off-wafer pulse communication of the BrainScaleS wafer-scale system.
A custom Kintex7 FPGA board serves 8 HICANN units on the silicon wafer, thus supporting the pulse transmission of up to 4096 neurons (both stimuli and hardware neurons). 
The characterization covers the playback module, the downstream and upstream path across 8 HICANN channels, and the trace module.

\subsubsection{Providing spike stimuli}
In the following we summarize key results for the downstream transmission. We always provide the technical results as well as equivalent biological numbers assuming the \num{10000} acceleration.
\begin{itemize}
    \item For 1 HICANN, the \emph{throughput} of \SI{17.8}{MEvent/s} (\SI{1.78}{\kilo\hertz} biological spike rate) is limited by the minimum time between the transmission of two pulse packets of \SI{56}{\ns} (\SI{0.56}{\ms}).
    \item The \emph{transmission delay} (difference between release in FPGA and release in HICANN) amounts to \SI{230}{\ns} (\SI{2.3}{\ms}). Note that this delay is considered when converting input spike times in \si{\ms} into FPGA release times in \si{\ns} when running SNN simulations defined in PyNN \cite{muller2022operating} such that the actual value has no effect on the results.
    \item The \emph{temporal resolution} for input stimuli is defined by the \SI{8}{\ns} clock in the FPGA (\SI{0.08}{\ms} biological time). Yet, the release of subsequent spikes in the HICANN happens at least \SI{56}{\ns} (\SI{0.56}{\ms}) after the previous spike, which may lead to an additional delay. Especially for Poisson spike trains with a biological firing rate $>\SI{1}{\kHz}$ this leads to jitter and a higher average transmission delay (Fig.~\ref{fig_delay_and_jitter}).
    \item \emph{Pulse loss} occurs when approaching the one-HICANN throughput limit. Even for Poisson spike trains it only starts at $\approx\SI{1.6}{\kHz}$ biological rate.
    \item When input pulses are sent to all 8 HICANNs, the \emph{overall throughput} is limited by the playback throughput of \SI{121}{MEvent/s} (\SI{12.1}{\kHz}), which is close to $8\times$ the per-HICANN throughput of \SI{143}{MEvent/s} (\SI{14.3}{\kHz}).
    \item Additional \emph{jitter} may occur when all 8 HICANNs are stimulated due to the packing of spikes into pulse groups in the FPGA.
\end{itemize}

The main limitations of the current design are as follows:
As the \emph{timestamp mode} in the FPGA-HICANN channels (cf.~Sec.~\ref{layer2}) was not used due to bugs in the heap sort unit \cite{scholze2010mixdes} of the current design, no double pulse packets are used for the transmission to the HICANNs. By fixing the heap sort, which was designed to sort pulse packets originating from other FPGAs or wafers, or replacing it with a simpler block supporting double pulse packets, one could improve the throughput to \SI{25}{MEvent/s} (\SI{2.5}{\kHz}) per HICANN.
At the same time or in addition, one could change the frame format of the playback memory (cf.~Fig.~\ref{playb_frame}): Instead of storing the 15-bit HICANN release timestamp (only used in \emph{timestamp mode}, hence not used here), one could store another 14-bit label per 32-bit word, allowing to potentially double the throughput of the playback module.
In addition, one should consider increasing buffer sizes to avoid pulse loss and adding counters to track pulse loss.

\subsubsection{Spike Recording}
The upstream spike transmission and recording to the trace memory was measured using regular and pseudo-random spike trains on the HICANNs.
\begin{itemize}
    \item For one HICANN the \emph{throughput} saturates at \SI{25}{MEvent/s} (\SI{2.5}{\kilo\hertz} biological spike rate) which corresponds to the physical limit of the FPGA-HICANN LVDS lines.
    \item \emph{Pulse loss} starts at \SI{18}{MEvent/s} (\SI{1.8}{\kilo\hertz}) and is more pronounced for pseudo-random spike trains (Fig.~\ref{fig_bg_loss}). Note that the pulse packet drop occurs in the HICANN's merger tree \cite{schmidt2023clean} due to back pressure and not in the FPGA.
    \item The trace module is able to serve all events from 8 HICANNs, such that upstream throughput is equal to $8\times$ the one-HICANN throughput: \SI{200}{MEvent/s} (\SI{20.0}{\kilo\hertz}).
\end{itemize}
The upstream transmission and the trace module are well matched to the physical limits of the FPGA-HICANN links. During real experiments one should make use of pulse drop counters in the HICANN to know whether spikes are lost.
To avoid pulse loss, one can also reduce the number of neurons mapped per HICANN, thus decreasing the actual pulse generation rate.

\subsection{Real Applications on BrainScaleS Wafer}
When the measurements for this paper were performed in 2016, it was not possible to run more complex SNN models on the wafer due to missing robust handling of faulty components and neuron calibration.
In the following years, several models have been realized of increasing complexity and size. 
\cite{schmitt2017neuromorphic} implemented a deep SNN for handwritten image classification on the wafer. For this, an equivalent DNN was first trained in software on a host computer and then converted to a SNN and mapped to the BrainScaleS hardware. The weights of the SNN were further tuned by an in-the-loop training where the spike trains of the neurons were recorded and treated as activations when applying the error backpropagation on the host, this way achieving a classification accuracy of \SI{95}{\%} getting close to the DNN (\SI{97}{\%}).
Another work \cite{kungl2019accelerated} implemented a SNN to run Bayesian inference (sampling from probability distributions), again applying an in-the-loop algorithm for training the weights.
While both models contained less than 1000 neurons, they used the presented FPGA-based Layer-2 network for input stimuli and spike recording. The measured downstream throughput per FPGA or HICANN are considered during the placement of input spike trains in the mapping software \texttt{marocco} \cite{muller2022operating}.

\cite{schmidt2023clean} shows how challenging it was to bring the BrainScaleS wafer-scale into operation given its complexity and issues due to faulty components in the mixed-signal circuits.
Finally, they demonstrated a chain of neuron populations able to propagate spikes across a full wafer. 
Follow-up work by \cite{schmidt2025demonstrating} demonstrated the emulation of neuro-scientific benchmarks, namely a balanced random work \cite{brunel2000dynamics} and a small version of the cortical microcircuit model \cite{potjans2014cell}, the latter containing nearly 8000 neurons and 2.4 million synapses. That work proved two unique capabilities of BrainScaleS making use of the 10.000 acceleration: to perform fast parameter sweep and to run simulations lasting over 1 year of biological time. Again, the Layer-2 network characterized in this article was used.

\subsection{Application in STDP experiments}
Experimental studies have observed the biological process of spike timing dependent plasticity (STDP), 
where the weight of a synapse is regulated according to the relative timing when the presynaptic and postsynaptic
neuron fires. This mechanism plays an important
role in learning, behavior and memory and has met great interest for neuroscience and modelers for further study and understudying. The proposed design in this paper supports the performance of STDP experiments like the one proposed in \cite{Thorpe_08}. There, neurons are stimulated by continuous spike trains, which represent a spatio-temporal input pattern alternated with Poisson activity. The neurons from this process learn to fire if and only if the input pattern is presented again.

A follow-up work \cite{gilson2011} includes experiments with a network of 1000 stimuli neurons with mean rate at \SI{20}{\hertz}. 
The network requires a time window of \SI{2000}{\second} to learn a provided pattern.
This experiment can be easily realized by the presented BrainScaleS Layer-2 network taking into account the performed communication characterization. The input stimuli can be distributed over 16 FPGA-HICANN links such that each link handles approximately 64 input neurons.  This set of neurons per link has a total pulse rate of \SI{1280}{\hertz}, which is well below the maximum permitted bandwidth of \SI{1.8}{\kilo\hertz} per link and does not suffer from pulse loss (Fig.~\ref{fig_loss_figure}, Poisson case).
The accelerated computation of the hardware shrinks significantly the run time of the experiment from \SI{2000}{\second} to only \SI{0.2}{\second}. 
The overhead for pulse transmission between the host and the FPGA can be further reduced by splitting the input into several batches and using the loop mode of the Playback module. The distribution of the input to several batches is performed also at the software simulation of this model \cite{Thorpe_08, gilson2011}.

\subsection{Comparison with other state-of-the-art systems}
The AER protocol is with no doubt the most popular communication protocol within the neuromorphic 
community.
Other FPGA infrastructures for pulse AER pulse transmission have been developed in academia like~\cite{serrano09,berge07,fasnacht11,park2016hierarchical}.
Table~\ref{tab_comparison} shows a comparison between them and our implementation.
Similar to \cite{park2016hierarchical}, this work provides a very detailed communication characterization but also a benchmark application.
The BrainScaleS FPGA node is one of the most 
sophisticated and compact designs, implementing a complete dataflow chain from a 
host unit via FPGA, communication ICs (DNCs) towards  neuromorphic chips and backwards allowing 
for stimulation, monitoring and control. It is also suitable for full configuration of the system by a host unit.    
This is in contrast to other published FPGA solutions, which are
mainly restricted to pulse transmission~\cite{serrano09,berge07,fasnacht11,park2016hierarchical}. 
Another important asset of the BrainScaleS FPGA node over the others is a  large Playback/Trace   
storage level. Specially this property makes the presented communication system 
unique in a sense that it enables both standalone and long-term operation. 
The stimulation and tracing of neural pulses as well as the configuration is served by two independent datastreams to/from the neuromorphic chips, allowing for high throughput, verification flexibility and negligible data collision~\cite{Thanasoulis12}.

\begin{table*}[ht]
\caption{\label{tab_comparison}
Comparison of the BrainScaleS FPGA-based off-wafer communication with related work.
}
\small
\begin{tabular}{|p{1cm}|p{2.2cm}|p{1.0cm}|p{0.6cm}|p{1cm}|p{1.9 cm}|p{1cm}|p{1.3cm}|p{4.2cm}|}
\hline 
Ref. & \multicolumn{3}{c|}{Communication Characterization}& Memory & AER event & Configur.  & Monitoring & Other functionality \\
& Bandwidth & Latency & Jitter &  size &   & data & &\\
\hline
This work  & 125 / 250~Mevent/s (stim/trace) & 230 ns & 8 ns & 1GB & 32 bit, synch.& yes & yes & long-run and precise Stimulation/Tracing, loop-Stimul. mode\\
\hline
\cite{serrano09} & 25 Mevent/s  & 80 ms  & -- & 2MB & 16 bit, asynch. & no & yes & video transformation\\
\hline
\cite{berge07} & 41.66 Mevent/s &(90+64) ns  &  -- & -- & 16bit, synch. & no & no & optimization up to 125 Mevents/s\\
\hline
\cite{fasnacht11} & 66 Mevent/s & 0.8 $\mu$s & -- & 2 GB & 16 bit, asynch. & yes & no & probabil. address-event mapping\\
\hline
\cite{park2016hierarchical} &  36 Mevent/s & \SIrange{0}{450}{\mu\second} & -- & 0.5 GB & 32-bit, synchronous & no & no & scalable event routing through hierarchical AER routers \\
\hline
\end{tabular}
\end{table*}

In addition to our own comparison in Table~\ref{tab_comparison}, we refer to Young et al.~2019 \cite{young2019review} for a review of the spike communication systems of large-scale neuromorphic systems, including TrueNorth, Loihi, Neurogrid, Braindrop,  BrainScaleS, SpiNNaker, Darwin, and Dynap-SEL.

Also note further work in the BrainScaleS eco-system:
In \cite{thanasoulis2021delaybased}, we presented an extension of the current FPGA design adding an inter-FPGA routing module which enables configurable axonal delays as for example required in synfire chains \cite{kremkow2010functional} for the stable propagation of spiking activity. To enable multi-wafer experiments, Thommes et al.~2021 \cite{thommes2021brainscales} propose a concept using the EXTOLL networking technology, where the Kintex7 FPGAs are connected to Tourmalet network interface cards.
Later, in \cite{thommes2022demonstrating}, this EXTOLL-based spike communication was demonstrated to connect multiple BrainScaleS-2 chips \cite{pehle2022brainscales2}.
An alternative approach to interconnect 12 BrainScaleS-2 chip is shown in \cite{ilmberger2025brainscales2multichipsysteminterconnecting} using a Zynq-FPGA as Aggregator board. Again, in both multi-chip BrainScaleS-2 setups the same Kintex7 FPGA nodes from this work are used.

%% file: 07_Conclusion.tex
In this paper we provided a detailed analysis of the off-wafer pulse communication of the BrainScaleS neuromorphic system.
A comprehensive characterization of the spike transmission between the external FPGA node and the neuro-synaptic cores (HICANNs) on the wafer was performed yielding quantitative numbers for the throughput, delay, jitter and packet loss and various conditions (regular and Poisson spike trains).
The availability of these numbers is essential for the mapping of input spike source to FPGA-HICANN channels and also for the interpretation of the recorded spike data of the spiking neural networks emulated on the neuromorphic wafer.
This was demonstrated and evaluated in a benchmark application using a loopback mode of the HICANNs.
Overall, this work proves the suitability of the off-wafer communication design to support long-term and precise spike stimulation and recording for neural network emulations running \num{10000} times faster than in biology.